\documentclass{chi2008pubsformat-jcdl}
\begin{document}
\title{Towards a Model of Understanding Social Search}
\numberofauthors{2} 
\author{
\alignauthor Brynn M. Evans \\
       \affaddr{University of California, San Diego}\\
       \affaddr{La Jolla, CA}\\
       \email{bmevans@cogsci.ucsd.edu}
\alignauthor Ed H. Chi \\
       \affaddr{Palo Alto Research Center}\\
       \affaddr{Palo Alto, CA}\\
       \email{echi@parc.com}
}
\date{}
\maketitle
\begin{abstract}
Search engine researchers typically depict search as the solitary
activity of an individual searcher. In contrast, results from our
critical-incident survey of 150 users on Amazon's Mechanical Turk
service suggest that social interactions play an important role
throughout the search process. Our main contribution is that we have
integrated models from previous work in sensemaking and information
seeking behavior to present a canonical social model of user
activities before, during, and after search, suggesting where in the
search process even implicitly shared information may be valuable to
individual searchers.
\end{abstract}




\section{Introduction and Related Work}



Surprisingly, researchers have thought about navigating and browsing
for information as a single user activity, centered on
eliciting users' information needs and improving the relevance of
search results. For example, Choo, Detlor \& Turnbull
\cite{Choo-WebInfoSeeking99} discussed categories of search behaviors
and motivations in information seeking, but they overlooked the
role of other individuals in search.
On the other hand, library scientists \cite{Wilson-InfoNeeds}
have observed for some time that friends and colleagues may be valuable information
resources during search. Similarly, recent authors have
begun to recognize the prevalence and benefits of \emph{collaborative search}
\cite{Morris-CollabSearch, Twidale-BrowsingAsCollab}. 


However, in addition to explicit collaboration in joint
search tasks \cite{Morris-CollabSearch}, we believe
that even implicit social experiences could
improve the search process. Therefore, the general term ``social
search'' may more suitably describe information seeking and
sensemaking habits that make use of a range of possible social
interactions: including searches that utilize social and expertise
networks or that may be done in shared social workspaces. This notion
certainly encompasses collaborative co-located search, as well as
remote and asynchronous collaborative and collective search. Our focus
in this paper is to explore a model of social search that may offer
suggestions for supporting social interactions in the information
seeking process.

\section{Survey}
We conducted a survey
asking 150 users to describe their most recent search act using
Amazon's Mechanical Turk service. Our survey was
designed to resemble critical-incident reporting, in which users self-report 
events that occurred relatively recently
\cite{Flanagan-CriticalIncident}. 
We prompted users to report information about the \emph{context}
(``What were you doing just before you searched?'')
and \emph{purpose} (``What prompted you to perform the search?'') of
the selected incident, and how (or if) they interacted with other
individuals prior to and following the primary search act. Selected
survey questions are presented below:

\small
\begin{enumerate}
\item What kind of information were you searching for?
\item Did you talk with anyone before you searched?
\item What steps did you take to find this information? 
\item What did you do just after you searched?
\item If other people were nearby, were you interacting with them or
  were they influencing your search process?
\item After you found the information, did you share it with anyone?
\item If yes, how did you share the information?
\end{enumerate}

\normalsize
Users provided background information on their profession, job roles
(\mbox{Table \ref{fig:sample_details}}), and levels of job expertise
as rated on a 5-point Likert scale (\mbox{Table \ref{fig:timejobexperience}}).
Two-thirds of search acts occurred on the same day that
our survey was filled out; 19.3\% on the day before; and
17.3\% more than 2 days ago.  

\begin{table}[b!]
\small
\centering
\begin{tabular}{| c | c | c | c |}
  \hline
  \textbf{Profession} & \textbf{\%Users} 
& \textbf{Job Role}  & \textbf{\%Users} \\

  \hline
  Education & 9.3 
& Manager   & 19.3 \\

  Financial & 8.7 
& Assistant  & 18.7 \\

  Healthcare & 6.7 
& CEO/Director & 8.0 \\

  Govt. Agency & 6.0 
& Customer Support  & 7.3 \\

  Retail & 6.0 
& Teacher   & 6.0 \\


  \hline
\end{tabular}
\normalsize
\caption{The most frequently occurring professions and job roles
  reported by users in our sample.}
\label{fig:sample_details}
\end{table}

\begin{table}[b!]
\small
\centering
\begin{tabular}{| c | c | c | c |}
  \hline
  \textbf{Search Duration} & \textbf{\%Users} 
& \textbf{Job Expertise}  & \textbf{\%Users} \\

  \hline
  $<$ 5 minutes & 44.7 
  & 5   & 33.3 \\

  5--10 minutes & 23.3 
  & 4   & 35.3 \\

  10--20 minutes & 10.7 
  & 3   & 20.7 \\

  20--30 minutes & 13.3 
  & 2   & 7.3 \\

  $>$ 30 minutes & 8.0 
  & 1   & 3.3 \\
  \hline
\end{tabular}
\normalsize
\caption{Information reported about duration of the search act and
  level of job experience.}
\label{fig:timejobexperience}
\end{table}



\section{Results}
\begin{figure*}[!htp]
\centering
\includegraphics[scale= 0.7]{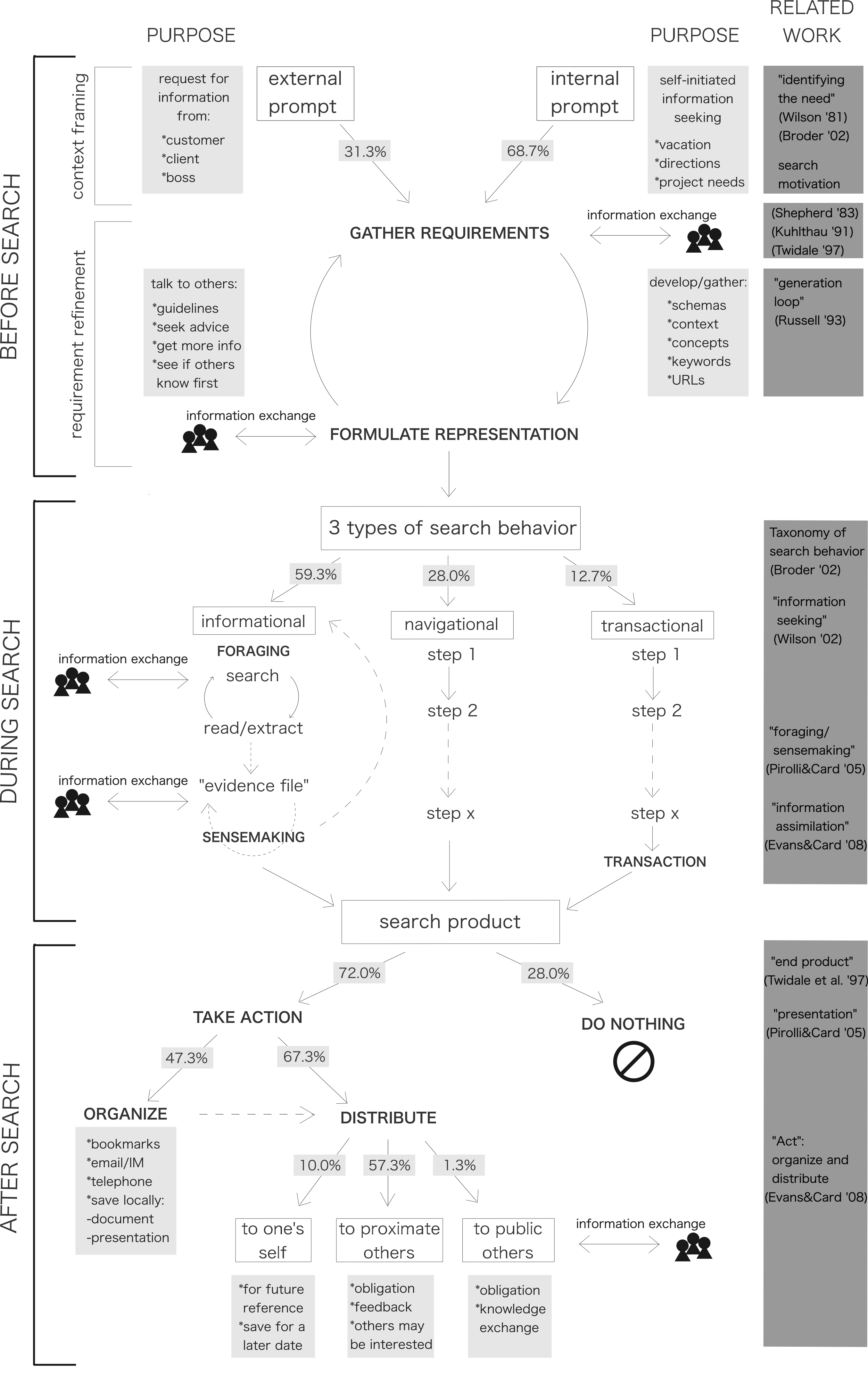}
\caption{Canonical social model of user activities before, during,
  and after a search act, including citations from related work in
  information seeking and sensemaking behavior.}
\label{fig:search-model}
\end{figure*}

Our main contribution is that we have integrated
our findings with models of sensemaking and information seeking from
the literature, and we present a canonical model of social search (Figure
\ref{fig:search-model} on the next page). We will discuss our model in three
phases, highlighting where information exchange occurred through social interactions and
providing both quantitative data and anecdotal case studies of actual
user behavior.    


\subsection{Before Search}
\subsubsection{Context Framing} 
Information-seeking behavior is rooted in a ``need'' to find
information \cite{Broder-Taxonomy, Wilson-InfoNeeds} or a motivation
that drives the search process. This may be thought of as the
\emph{context framing} stage,
where user motives and information needs are defined. Requests for
information may come from an external source or may be
self-initiated. From our sample, 47 of 150 users (31.3\%) searched
for information following a specific request from a boss, customer, or
client; whereas 103/150 users (68.7\%) were self-motivated to find
information related to personal or work endeavors.

\subsubsection{Requirement Refinement} 
After the information need and motives are established, users \emph{refine
their search requirements}. Previously described as a \emph{generation
  loop} \cite{Russell-Sensemaking}, this phase involves gathering
requirements and formulating relevant schemas such that an effective
search may result. In our data, this cycle is marked by social
interactions 42.0\% of the time (63/150 users) as a means to ``influence the
information need'' \cite{Twidale-BrowsingAsCollab}, gather useful
keywords or URLs, or develop the necessary guidelines for a given task. 


\subsection{During Search}
Although search can be a cyclical process, the search stage
in our model represents the active instantiation of representations or
``encodons,'' as part of a ``data coverage'' loop
\cite{Russell-Sensemaking}. In other words, this is the stage where
users engage in traditional information seeking
\cite{Wilson-InfoNeeds} and foraging activities
\cite{PirolliCard-InfoForaging, PirolliCard-Sensemaking05}. We
detail each of the three types of search acts below (\emph{transactional,
navigational}, and \emph{informational} \cite{Broder-Taxonomy}), drawing
special attention to the social interactions.

\subsubsection{Transactional Search}
With a \emph{transactional} search, users locate a source where they can
subsequently perform a transaction or other ``web-mediated activity''
\cite{Broder-Taxonomy}. In our sample, this typically involved
navigating to a website through a series of routine steps and
requesting specific information such as weather at a destination,
movie listings, or data from a customer's account. As an example, an
ambulance chief found the distance from a patient's home to the
hospital through MapQuest.com by entering the start and end
locations, and then retrieving the mileage information.

\subsubsection{Navigational Search}
During a \emph{navigational} search, users perform a series of
actions to identify content from a particular, often familiar,
location. The content is often known in advance, or will be easily
recognized once it is (re)discovered. For example, a nurse used the
NIH website to look up information on a medical drug: first by logging
into Google, looking up the NIH's web address, and then searching for
the drug on the NIH website. The nurse reports: ``I knew exactly where
[the information] would be---just couldn't recall what the answer was.''  

\emph{Transactional} and \emph{navigational} searches occasionally
involved pre-search interactions with others (42.1\% and 47.6\%,
respectively), but information exchange did not occur during the
search itself. Consequently, it is unlikely that socially-augmented search
would improve or facilitate \emph{transactional} or
\emph{navigational} information retrieval.

\subsubsection{Informational Search}
On the other hand, social search may greatly improve tasks involving
\emph{informational} search---an exploratory process, combining
foraging and sensemaking \cite{PirolliCard-InfoForaging,
  PirolliCard-Sensemaking05}, of searching for information that may or
may not be familiar to the user.

\textbf{Foraging.} The basic ``information assimilation'' process
\cite{EvansCard-CHI08} illustrates
this early foraging phase where users search for information within a
specific patch, followed by skimming, reading, and extracting
information from source files. Throughout this process, users may
update and shift their search representations
\cite{Russell-Sensemaking} as they discover new items, at times by seeking  
feedback from others. A public librarian worked
with her boss to find the Cheetah \mbox{Girls 2} soundtrack: ``We had
to deduce a number of [keyword] combinations. We tried a
number of ways to write Cheetah Girls, including hyphens and spelling
out the number two.''


\textbf{Sensemaking.} After an initial pass at foraging, users may identify 
preliminary ``evidence files'' \cite{PirolliCard-Sensemaking05} from
which they can further modify their search
schema and query. For example, an English Professor engaged in a
classic sensemaking process when he used MSN.com to look up information
about Robert Frost. He copied and pasted query results into a Word
document, which he revisited later to sort and summarize for an
upcoming lecture.

Social interactions may also augment the sensemaking process. 
One programmer from Intuit engaged with a colleague while
searching for an application programming interface (API). They began
with a brainstorming session, followed by an online search, and finally
``another round of discussion'' on whether the API (the
``evidence file'') would be sufficient for their purposes. 

Over half of the search experiences reported by our sample were
\emph{informational} in nature (89/150 users, 59.3\%). Even though our survey did
not ask users to report social incidents or explicitly collaborative
search acts, searchers did engage with others both before and during
the \emph{informational} search. In fact, 35 out of these 89 individuals
(39.3\%) had social experiences prior to searching---and not simply
out of obligation. Some used these social opportunities for
brainstorming, to assess others' opinions, or to improve their own
search schemas (``to know what kind of material would be useful'').
Consequently, there appears to be both a need for and an interest in input
from others throughout the \emph{informational} search process.

\subsection{After Search}
Following the active search phase, an ``end product'' is often
obtained \cite{Twidale-BrowsingAsCollab}, which may be ``acted'' on
through \emph{organization} and/or \mbox{\emph{distribution}
  \cite{EvansCard-CHI08}}. 

\subsubsection{Organization}
Nearly half of our users (71/150 or 47.3\%) organized their end
products in some fashion. Pirolli and Card
referred to this process as 
\emph{schematizing}, where raw evidence is organized and
``represented in some schematic way''
\cite{PirolliCard-Sensemaking05}.
For example, one real estate agent printed and reviewed
the results of a search before ``giving them to an attorney for legal
inspection.'' The president of a design company bookmarked online
articles about web mashups to read later in the week. 
Evans and Card remarked how users
may serve as filters for information through their actions in
bookmarking, tagging, or annotating items, which subsequently
``improve[s] access to the item in the future or...integrate[s] it
with previous knowledge and \mbox{context'' \cite{EvansCard-CHI08}}. 

\subsubsection{Distribution}
In fact, most of these organizational acts served as a means for
distributing the end product to others, as a ``presentation or
publication of a case'' \cite{PirolliCard-Sensemaking05}.  
At other times, information was shared directly face-to-face or
verbally over the phone. Regardless of the
distribution mechanism, over half of respondents (88/150, 58.7\%) shared their
end products with others (e.g., a floral designer relayed information
about local spring blooming flowers to a bride-to-be). Some even
``shared'' the content with themselves by printing out documents or
bookmarking websites for re-accessing at a future date (11 users,
7.3\%).

These activities suggest that social
interactions are important even after the primary search act,
especially for self-initiated searches. 49 of the 104 self-motivated
searchers (47.5\%) distributed search content for verification, feedback, or
because they thought others would find it interesting.

\section{Conclusion}
As we outlined through the model, social inputs may help users
throughout the search process. Before searching, social
interactions may help establish the requirements for the actual
search task. During search, especially for self-motivated
\emph{informational} searches, users may talk to others for advice,
feedback, and brainstorming to improve their search schema and query
keyword selections. After search, users may still wish to engage with
others to collect additional feedback or to share knowledge gained
during the search. 

Altogether this suggests that a notion of ``social search'' may
facilitate the process of information seeking. 
One techinque for supporting social search may be through instant
messaging access to your personal connections alongside the search
box. Or it might exploit a website's existing community to reveal
domain-specific experts who would be willing to advise
searchers. Alternatively, sites could display related and successful
keyword combinations or search trails from previous users, or
automatically-generated tag clouds of semantically related concepts
that may provide high-level feedback on the general search topic. We
hope that our discussion in this paper will encourage
researchers to explore both explicitly collaborative social interactions
as well as implicitly shared information to augment web-based search.


\bibliographystyle{abbrv}

\end{document}